\documentclass[aps,epsfig]{revtex4}
\usepackage{graphicx}
\begin{document}
\def\be{\begin{equation}}
\def\ee{\end{equation}} 
\def\bfi{\begin{figure}}
\def\efi{\end{figure}}
\def\bea{\begin{eqnarray}}
\def\eea{\end{eqnarray}}

\title{Complex phase-ordering of the one-dimensional Heisenberg 
model with conserved order parameter}

\author{R. Burioni}
\affiliation{Dipartimento di Fisica and INFN, Universit\`a di Parma, 
Parco Area delle Scienze 7/A, I-423100 Parma, Italy}
\author{F.Corberi}
\affiliation{Dipartimento di Matematica ed Informatica
via Ponte don Melillo, Universit\`a di Salerno, 84084 Fisciano (SA), Italy}
\author{A. Vezzani}
\affiliation{CNR-INFM and Dipartimento di Fisica, Universit\`a di Parma, 
Parco Area delle Scienze 7/A, I-423100 Parma, Italy.}

\begin{abstract}
We study the phase-ordering kinetics 
of the one-dimensional Heisenberg model with conserved order parameter, 
by means of scaling arguments and numerical simulations.
We find a rich dynamical pattern with a
regime characterized by two distinct growing lengths.
Spins are found to be coplanar over regions of a typical size
$L_V(t)$, while inside these regions smooth rotations
associated to a smaller length $L_C(t)$ are observed. 
Two different and coexisting ordering mechanisms
are associated to these lengths, leading to
different growth laws $L_V(t)\sim t^{1/3}$ and $L_C(t)\sim t^{1/4}$
violating dynamical scaling. 
\end{abstract}

\maketitle


\section{Introduction} \label{intro}

After quenching a ferromagnetic system to a low temperature phase, 
relaxation towards the 
new equilibrium state is realized by a progressive phase-ordering~\cite{Bray94}.
The specific mechanisms involved in the coarsening phenomenon
depend on the presence and the nature of topological defects.
In $d$-dimensional systems described by and ${\cal O}(N)$ vector
order parameter, topological defects are unstable 
for $N>d+1 $ (for $N=d+1$ peculiar defects as textures~\cite{Rutenberg95} may
be present). Therefore, in the asymptotic regime when
all defects have disappeared, 
the dynamics is solely driven by the reduction 
of the excess energy related to the smooth rotations 
of the order parameter. In contrast, 
systems with $N\le d$ are characterized by
the presence of stable defects whose presence influences
the dynamics in the whole phase-ordering stage. In
particular, when $N=d$ defects are localized and 
ordering occurs by
mutual defect-antidefect annihilation.
This is the case of the Ising chain, where 
up and down domains are separated by point-like
interfaces performing random walks.

Generally, the late stage is characterized by
dynamical scaling \cite{Bray94,Furukawa84}. This implies that 
a single characteristic length
$L(t)$ can be associated to the development of order 
in such a way that configurations of
the system are statistically independent of time when
lengths are measured in units of $L(t)$.  
The characteristic length usually has a power law growth 
$L(t)\propto t^{1/z}$. 
In systems with a conserved order parameter (COP) one
generally finds $z=3$ \cite{Huse86,Bray89} or 
$z=4$ \cite{Corberi98,Bray89} for $N=1$ and $N>1$ respectively. 

For systems at or below the lower critical dimension $d_L$,
such as the Ising chain, a true asymptotic phase-ordering 
can only be observed in quenches to $T=0$. However,
if quenches to a relatively low temperature are performed,
one observes an initial transient regime (but very long lasting
when $T$ is small) where the dynamics is indistinguishable from
that at T=0. This regime lasts until
$L(t)$ has grown comparable to the
equilibrium coherence length $\xi (T)$. 
 
In this Article, we investigate the phase-ordering
kinetics of the one-dimensional Heisenberg model ($N=3$)
with COP quenched to a
low temperature $T$. We show that the dynamics is much
richer than what one would naively expect. 
This is due to the formation, in an early stage,
of couples of parallel spins, separating regions in which the
spins are coplanar. These parallel spins
act as point-like defects, as it 
will be explained in Sec.~\ref{num}. Their presence
provides an analogy between the Heisenberg 
and the Ising chain, where couples of parallel spins
and regions of coplanarity
in the former model correspond to interfaces and domains
in the latter. The analogy is not only formal, but is
reflected in the kinetics: in a first stage
(whose duration however diverges in the $T\to 0$ limit)
regions of coplanar spins coarsen 
similarly to the domains of the Ising model, 
their typical length growing as
\be
L_V(t)\sim t^{1/3}.
\label{lv}
\ee
In this regime the number of defects is reduced only
by a mechanism which recalls the annihilation of the interfaces in 
the Ising model.
Simultaneously, smooth rotations
of the spins, typical of vectorial systems, occur inside the
regions of coplanarity.
The coherence of the spins inside these regions extends over
a length $L_C(t)\ll L_V(t)$ increasing as
\be
L_C(t)\sim t^{1/4}.
\label{lc}
\ee 
The existence of two growing length, associated to
different ordering mechanisms, produces
the breakdown of dynamical scaling. 
In the analogy between the Heisenberg and the Ising chain, a notable
difference must be stressed. While in the latter interfaces are
stable defects which can only be removed by mutual annihilation,
defects in the former are unstable. Namely, after a first stage
of coarsening of coplanar regions, the defects spontaneously
decay due to thermal fluctuations. The typical lifetime
of the defects being limited by temperature,
coarsening of the coplanar regions persists up to very
long times in deep quenches. 
After that, defects disappear and the
system finally enters a late stage where smooth spin rotations
remains the only mechanism at work, until
equilibration is attained when $L_C(t)\sim \xi (T)$. 
In this regime dynamical scaling is restored
with exponent $z=4$, as expected for a system with $N>d+1$.  
In the small temperature limit the duration of the two regimes, with 
and without dynamical scaling, are comparable.

This paper is organized as follows:
In Sec.~\ref{model} we introduce the model and define the
observable quantities that will be considered.
In Sec.~\ref{num} we describe the main features of the dynamics in
the different regimes, compute the value of the exponents
and of other quantities by means of scaling arguments,
and compare our results with the outcome of numerical
simulations of the model.
A summary and the conclusions are contained in Sec.~\ref{concl}.

\section{Model and observables} \label{model}

The Heisenberg model is defined by the Hamiltonian
\be
H[\sigma ]=\sum _{i=1}^{\cal N}\epsilon _i=
-J\sum _{i=1}^{\cal N}(\vec \sigma _i \cdot \vec \sigma _{i+1}-1)=
-J\sum _{i=1}^{\cal N}(\cos \phi _i-1),
\label{hamiltonian}
\ee
where $\epsilon _i$ is the local energy density, $\vec \sigma _i$ is a three-components unit vector spin, $i=1,...,{\cal N}$
are the sites on a one-dimensional lattice and $\phi _i$ is the angle
between $\vec \sigma _i$ and $\vec \sigma _{i+1}$.
We will assume $J=1$ and the Boltzmann constant $k_B=1$. 

The equilibrium properties of the model are exactly known~\cite{Fisher63}.
This system is ergodic except at $T=0$.
At any finite temperature the state is disordered
with a vanishing magnetization and internal energy (per spin) 
$E_{eq}(T)=T-\coth (1/T)+1$ with the low temperature expansion 
$E_{eq}(T)\simeq T$. The correlation function 
$C_{eq}(r)=\langle \vec \sigma _i \vec \sigma _{i+r}\rangle=[1-E_{eq}(T)]^r$
decays exponentially over a coherence length $\xi (T)$ that
diverges in the $T\to 0$ limit.

Concerning dynamics, at equilibrium the model possesses an intrinsic
kinetics where the energy and the magnetization are conserved \cite{intrinsic}.
Studies \cite {scalbreak} of this intrinsic dynamics have evidenced
the failure of the equilibrium scaling symmetry close to $T=0$.
Experimental work supports this picture \cite{expbreak}.

\begin{figure}
    \centering
    
   \rotatebox{0}{\resizebox{.85\textwidth}{!}{\includegraphics{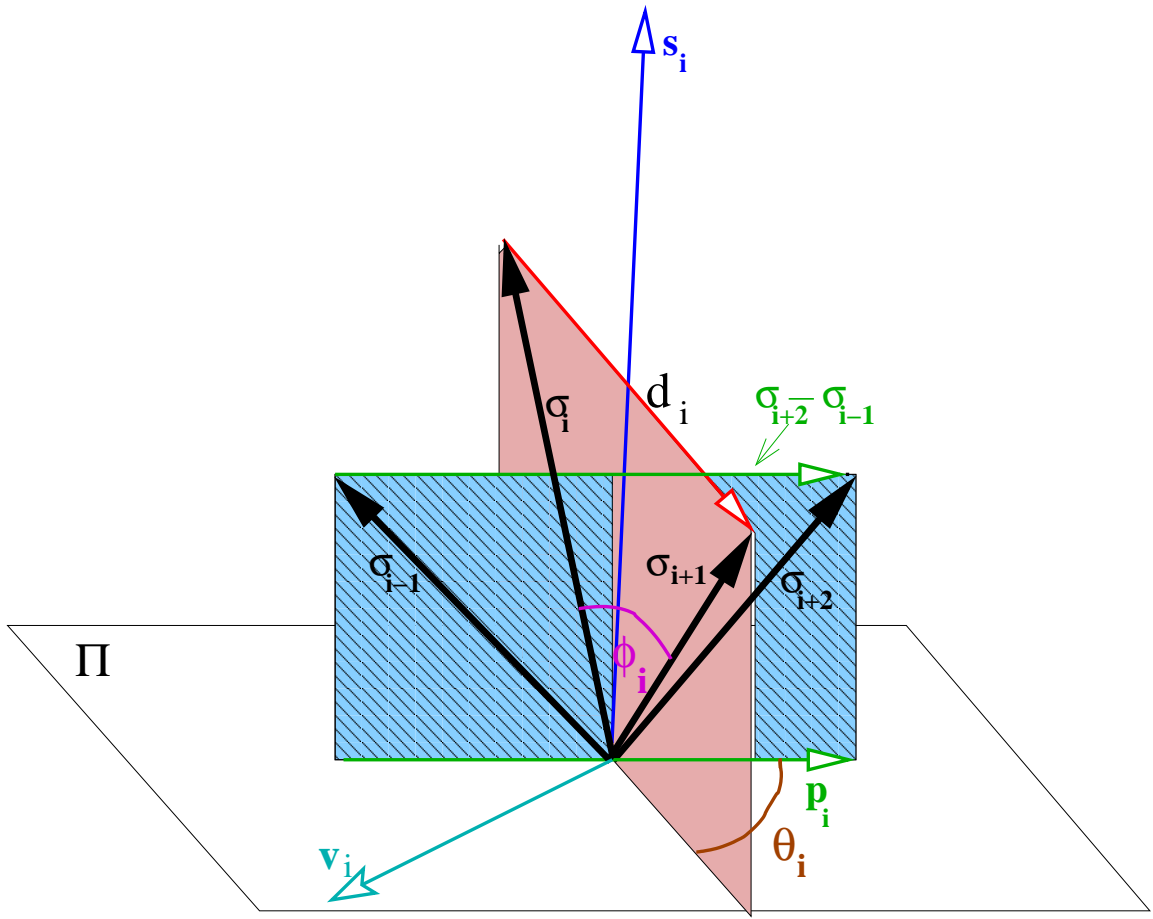}}}
    \caption{Schematic representation of the four spins 
$\vec \sigma _{i-1},\vec \sigma _{i},\vec \sigma _{i+1},\vec \sigma _{i+2}$.
}
\label{move}
\end{figure}

In this paper we consider a different situation, where the system is
quenched from an high temperature configuration to a low temperature
$T$, transferring the excess energy to a heat bath coupled to it.
In order to model this, we consider a dynamics where two neighboring spins
$\vec \sigma _i,\vec \sigma _{i+1}$ of a configuration $[\vec \sigma ]$ are randomly chosen 
at each time-step and then they
are updated to $\vec \sigma '_i,\vec \sigma '_{i+1}$
provided the local magnetization is conserved, namely 
$\vec s_i=\vec \sigma _i+\vec \sigma _{i+1}=\vec \sigma '_i+\vec \sigma '_{i+1}$.
Notice that magnetization is conserved at the local level with this rule.  
Due to the conservation law, the spins 
$\vec \sigma _i,\vec \sigma _{i+1}$
can only rotate rigidly around their sum $\vec s_i$,
as shown in Fig.~\ref{move}. 
We consider the heat-bath transition rates  
$[\vec \sigma ]\to [\vec \sigma ']$ satisfying detailed balance
\be
w_i[\vec \sigma']=W_i^{-1}
\exp \left (-\frac {H[\vec \sigma ']}{T}\right ),
\label{heatb1} 
\ee
where $W_i=\int d\vec \sigma_{i} ' d\vec \sigma_{i+1} ' \delta(\vec \sigma_{i}+\vec \sigma_{i+1}-\vec s_i) 
\exp(-{H[\vec \sigma ']}/{T})$. Heat-bath transition rates provide
a particularly fast and efficient dynamics with respect to other (i.e. Metropolis) 
choices \cite{notadin}. 
Let us denote $ \vec d_i = \vec \sigma_{i+1}-\vec \sigma_i $, 
$ \vec d'_i =\vec \sigma'_{i+1}-\vec \sigma'_i $ and 
$\vec p_i$ the projection of $\vec \sigma _{i+2}-\vec \sigma _{i-1}$ on the plane $\Pi $ 
perpendicular to $\vec s_i$ (see Fig.~\ref{move}).
Any move, involving  the couple $\vec \sigma _i$ and $\vec \sigma _{i+1}$,
can be described by a rotation in the plane $\Pi$  from $\vec d_i$ to $\vec d_i'$. In this framework 
the angles between $\vec p_i$ and $\vec d_i$ ($ \vec d_i'$), denoted as $\theta_i$  
($\theta'_i$), fully parametrize the dynamics and the 
transition rate (\ref{heatb1}) can be rewritten as
\be
w_i(\theta'_i)=W_i^{-1}
\exp \left (\frac {d_i p_i (\cos \theta'_i -1)}{2T}\right ),
\label{heatb2} 
\ee 
where 
$d_i=|\vec d_i|=|\vec d_i'|$, $p_i=|\vec p_i|$.
Von Neuman rejection method \cite{knuth}  allows to efficiently generate
$\theta_i'$ according to the transition rates (\ref{heatb2}).
Notice that in a move  the typical deviations from the lowest energy configuration
($\theta _i'=0$) are of order 
\be
\cos \theta _i'-1\sim \frac{2T}{d_ip_i}.
\label{typictheta}
\ee

We consider a system initially prepared in an
high temperature uncorrelated state, with $\sum _{i=1}^{\cal N}\vec \sigma  _i=0 $, and then quenched, at time $t=0$,
to a lower final temperature $T$. 
As already mentioned in the Introduction, 
the dynamics of systems at or below the critical dimension, such as the
one considered here, is characterized 
by an initial transient where
the system orders \cite{notacrit} as in a quench to $T=0$.
The characteristic length of ordered regions grows in time until, 
at time $\tau _{eq}(T)$ it becomes
comparable to $\xi (T)$. At this point the final equilibrium state
at $T$ is entered and phase-ordering ends. 
If the system is quenched to a 
sufficiently low temperature, since $\xi (T)$ is very large, 
the phase-ordering kinetics extends over a huge time window 
$t<\tau _{eq}(T)$. 

Characteristic lengths, and scaling properties, can be studied from the
knowledge of the two-points equal time correlation function
\be
C(r,t)=\langle \vec \sigma_i(t) \cdot \vec \sigma_{i+r}(t) \rangle,
\label{gdir}
\ee
where $\langle \dots \rangle$
means an ensemble average, namely taken over different initial conditions
and thermal histories.
Due to space homogeneity, $C(r,t)$ does
not depend on $i$.
Dynamical scaling~\cite{Bray94} would imply
\be
C(r,t)=c(x),
\label{scalgferro}
\ee  
where $x=r/L(t)$.
For systems with with an ${\cal O}(N)$ vector order parameter the
Bray-Puri-Toyoki~\cite{BPT} behavior 
\be
1-c(x)\sim x^N
\label{brayputo}
\ee
is observed for small $x$.
In the scalar case, this behavior reduces to the Porod´s law~\cite{Porod} 
\be
1-c(x)\sim x,
\label{porod}
\ee
which is generally expected in systems with sharp interfaces.
From Eq.~(\ref{scalgferro}) one can extract a quantity $L_C(t)$ proportional to $L(t)$
from the condition
\be
C\left [L_C(t),t\right ]=\frac{1}{2},
\label{halfheight}
\ee
namely as the half-height width of $C(r,t)$. 
In the following we will also consider the correlation 
\be
V(r,t)=\langle \vec v_i(t) \cdot \vec v_{i+r}(t) \rangle.
\label{vdir}
\ee 
where the unit vectors
\be
\vec v_i=\frac {\vec \sigma _i \times \vec \sigma _{i+1}}
{\vert \vec \sigma _i \times \vec \sigma _{i+1}\vert }
\ee
identify the planes formed by neighboring spins, hence $V(r.t)$ represents the correlations 
between these planes. 
When scaling holds, $V(r,t)$ behaves similarly to $C(r,t)$, 
namely
\be
V(r,t)=v(x),
\label{vscal}
\ee  
and, defining $L_V(t)$ analogously to $L_C(t)$ in Eq.(~\ref{halfheight}), also $L_V(t)\propto L(t)$.

\section{The dynamics: Scaling arguments and numerical results}\label{num}

In the following we will discuss the main features of the dynamical
process, by means of scaling arguments and numerical simulations. 
These are performed on a string of  
$8000$ spins with periodic boundary conditions $\sigma _{{\cal N}+1}=\sigma _1$. We have checked that with this size our
simulations are free from finite size effects. 
An average over $5-10$ realizations is made for each simulation. 
In the limit of low temperatures, to
which we are interested in, the kinetics becomes very slow since, as it will
be discussed below, time rescales as $T^{-1/2}$; simulations are therefore quite time consuming in this region. 

The dynamics of a low temperature quench is characterized by
different subsequent regimes, which are separately discussed below. 

\subsection{Pinning (Quenches to $T=0$).} \label{pinning}

\begin{figure}
    \centering
    
   \rotatebox{0}{\resizebox{.85\textwidth}{!}{\includegraphics{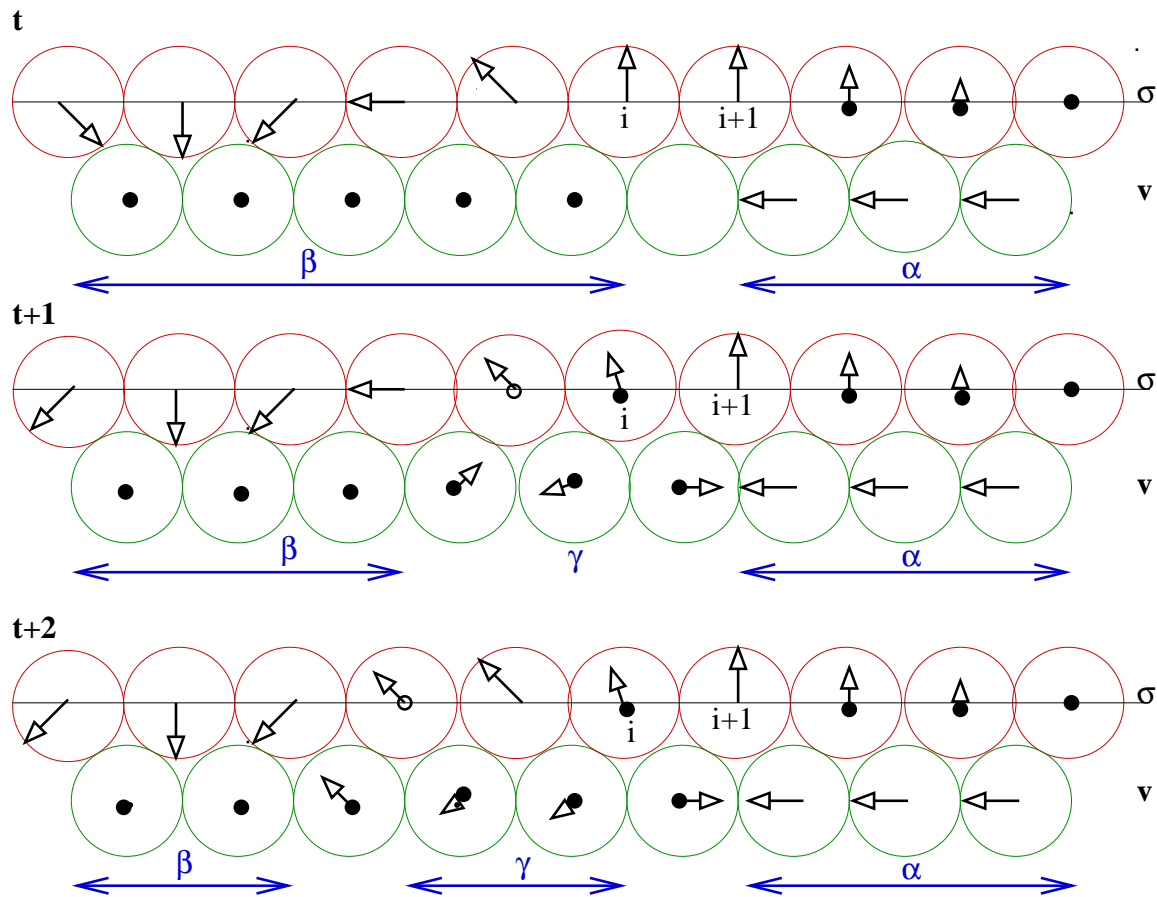}}}
    \caption{Schematic representation of spin configurations at three 
subsequent times $t,t+1,t+2$. For each time the upper and lower lines represent the 
$\vec \sigma $ and $\vec v$ configuration.
Given a vector $\vec \sigma $ (or $\vec v$) with components $(a,b,c)$,
the arrow in the figure is a vector of components $(a,b,0)$, namely the
projection of $\vec \sigma $ on the two-dimensional plane of the figure. 
The third component $c$ can be read off by the constraint of unitary length
of $\vec \sigma $, with the help of the unitary circles represented around
each vector. The origin of a vector is marked with an heavy dot when $c<0$
(vector pointing behind the figure), or with an open circle
when $c>0$. The meaning of the {\it phases} $\alpha,\beta,\gamma$
is discussed in the text.}
\label{figure}
\end{figure}

The kinetics of the quench to $T=0$ is determined by the
existence of frozen states where the system gets trapped after a while. 
The nature of these states can
be understood by looking at the first line (denoted as time $t$)
of Figure~\ref{figure}. The two 
spins on sites $i, i+1$ are parallel, hence no move involving
this couple can be done, since the angle $\theta _i$ is not
defined. 
On the left and on the right of the aligned spins there are
regions where the spins are coplanar, and hence the $\vec v _i$
are parallel and point along certain directions, denoted 
by $\alpha, \beta$ etc., which can be considered
as different {\it phases} of the system. These phases
extend up to another couple of parallel spins (not shown in the figure).
As it will be shown in Sec.\ref{regime1}, in quenches to finite temperatures
the system depins after a while and these coplanar regions
coarsen much in the same way as equilibrium phases do
in usual coarsening systems. Due to this analogy,
the term phases is used also here.
However it must be noticed that regions of coplanarity
are not equilibrium phases, because in equilibrium
spins are coplanar AND aligned, while here they typically
rotate (see Fig.~\ref{figure}), as will be discussed in Sec.~\ref{regime1}.
For the
following discussions, we also introduce the
terminology of {\it distance} between two phases 
$\alpha, \beta$, related to the angle by which the vectors $\vec v_i$
of the phase $\alpha $ must rotate in order to align with those of
the phase $\beta $. In this sense we will also talk of
{\it nearby} or {\it distant} phases.

As anticipated in Sec.~\ref{intro}, 
the impossibility 
to eliminate couples of parallel spins by means of local 
rearrangements, involving $\vec \sigma _i,\vec \sigma _{i+1}$ alone,
make them reminiscent of topological defects.
Actually these spins represent real topological defects, as is readily
seen by considering the representation
in terms of $\vec v_i$ instead of $\vec \sigma _i$.
With this description, phases
are domains in a strict sense, namely $\vec v_i$ is constant
in the interior of coplanar regions. 
They will be denoted as {\it domains } in the following, or
planes, in view of the coplanarity of their spins, without
further specification.  
Different domains are separated by 
sharp boundaries and, right on top of them, there is a localized defect
where $\vec v_i$ is not defined. Then, once the proper representation
is considered, parallel spins qualify as defects in the
usual sense. 
Since spins are coplanar inside the domains and parallel
on a defect, it is clear
that any move involving any couple of spins is forbidden
at $T=0$, and the dynamics is frozen on states like those depicted
on the first line of Fig.~\ref{figure}.
By identifying aligned spins as defects
separating domains, an analogy with the COP Ising model, which freezes
as well at $T=0$~\cite{Cornell91}, can be drawn.
Let us mention, however, at least two main differences. First, 
the constraint on the motion of parallel spins in 
the Heisenberg model is
related to the kinetic rule and not
to $T=0$. Second, as will be discussed in Sec.~\ref{brakedown},
the defects in the Heisenberg model are unstable, although their
lifetime diverges in the $T\to 0$ limit.

\begin{figure}
    \centering
    
   \rotatebox{0}{\resizebox{.85\textwidth}{!}{\includegraphics{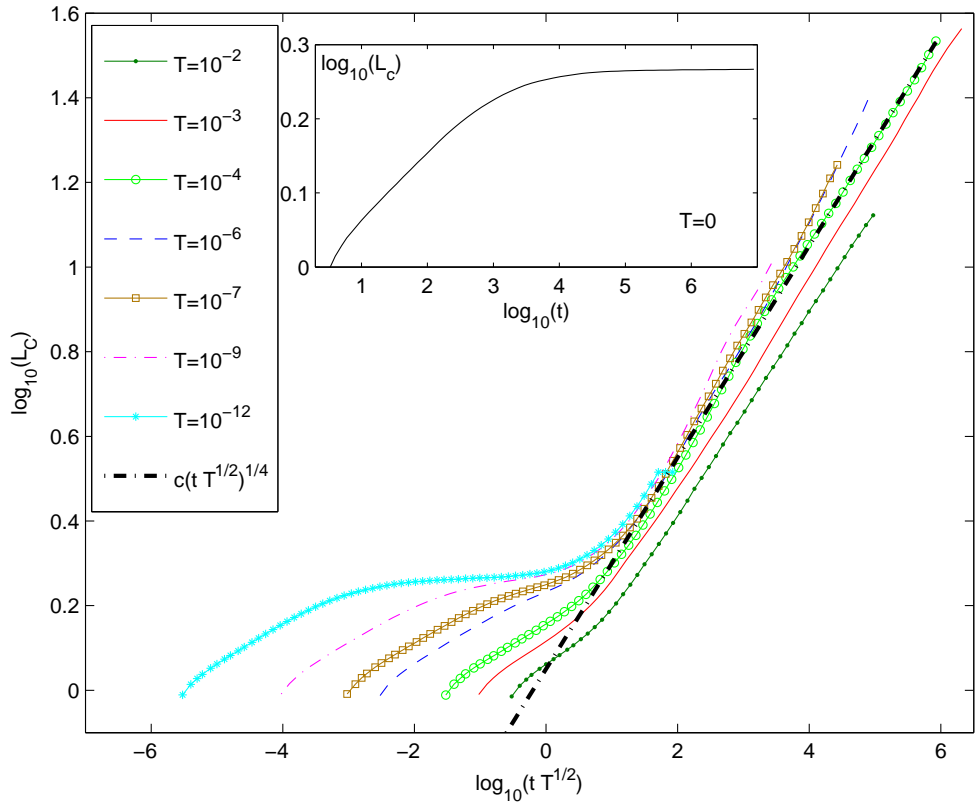}}}
\caption{The quantity $L_C(t)$ is plotted against rescaled time $tT^{1/2}$ for different temperatures. The dot-dashed line is the expected behavior $L_C(t)\propto (tT^{1/2})^{1/4}$
after the depinning (see text). In the inset the same
quantity is plotted against time in the case $T=0$.}
\label{lun_spin}
\end{figure}

\begin{figure}
    \centering
    
   \rotatebox{0}{\resizebox{.85\textwidth}{!}{\includegraphics{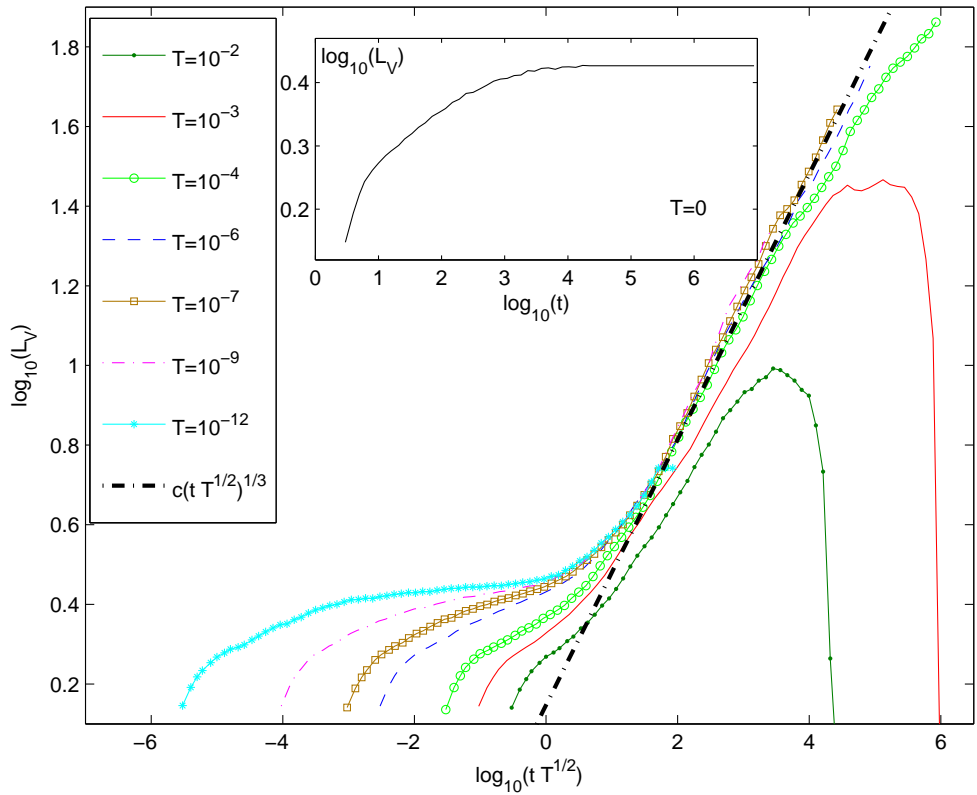}}}
\caption{The quantity $L_V(t)$ is plotted against rescaled time $tT^{1/2}$ for different temperatures. The dot-dashed line is the expected behavior $L_V(t)\propto (tT^{1/2})^{1/3}$
 in the first phase-ordering regime (see text). In the inset the same
quantity is plotted against time in the case $T=0$.}
\label{lun_piani}
\end{figure}

When a quench to $T=0$ is performed, the system starts reducing its energy
by ordering the spins until some couples happen to be nearly
parallel. In the meanwhile the $n$ spins between two defects adapt 
themselves on a plane. Since $n$ is a finite number this process
can be accomplished in a finite time.
At this point, the model gets trapped in one of the absorbing states
discussed above. 
Notice that, in a situation as the one discussed here, $L_V(t)$
and $L_C(t)$ describe, respectively, the length of
the domains and the coherence length of the spins
in the bulk of the planes.  
The evolution of the model toward the pinned state can be studied 
by following the evolution of these lengths in the insets
of Figs.~\ref{lun_spin},~\ref{lun_piani}. In a quench to 
$T=0$ both these quantities 
initially grow but then saturate to a constant value when the system freezes.    

\subsection{Depinning} \label{depinning}

When $T\neq 0$ but sufficiently small, the dynamics leading
the system to the frozen state proceeds practically as in the
case $T=0$ described in Sec.~\ref{pinning}, 
since it is entirely dominated by the moves which lower the energy
which are not affected much by a small $T$.
When the system is trapped in the absorbing state, however,
activated moves  
can occur inside the domains if $T>0$.
According to Eq.~(\ref{typictheta}), since the system is still
very disordered in this stage, both $d_i$ and $p_i$ are on average
large, and the typical value $\overline \theta$ of the  angle 
$\theta _i$ 
(we will use the over-bar in the following to denote the typical
value of a quantity) is small.
Notice, in fact, that even for the smallest temperatures considered 
in the simulations reported in Figs.~\ref{lun_spin},\ref{lun_piani}
pinning is never complete: $L_C(t)$ and $L_V(t)$ keep slowly
increasing because couples of spins are tiny twisted.
This mechanism eventually depins the system from
the absorbing state, restoring the dynamics, 
as shown in Figs.~\ref{lun_spin},\ref{lun_piani},
after a characteristic time $\tau _p(T)$.
From Eq.~(\ref{typictheta}), assuming $\overline \theta $ to be small, 
one has 
$\cos \overline \theta -1\simeq \overline \theta ^2\simeq 2T/(\overline d \overline p)$
and hence $\overline \theta \sim T^{1/2}$.
Since $\overline \theta $ is  tiny, a number 
$n(T)\simeq \pi /\overline \theta$ of these moves is needed in order
to produce an appreciable decorrelation 
(of an angle of order, say, $\pi $) with respect to the pinned state
and to restart the dynamics.
Therefore we find
\be
\tau _p(T)\sim T^{-1/2}.
\label{taup}
\ee
According to this result, for sufficiently low temperatures
the curves for $L_C(t)$ or $L_V(t)$ (and, more generally of
any time dependent observable), should collapse after the pinning
stage when plotted against $tT^{1/2}$. 
Figs.~\ref{lun_spin},~\ref{lun_piani} show that this is indeed 
quite well verified over 10 decades in temperature (with some 
deviations that will be discussed in Sec.~\ref{regime1}). 

\subsection{First phase-ordering regime: presence of domains}
\label{regime1}

Here we give a schematic description of the microscopic kinetics
in a first dynamical regime occurring after the depinning.
In this regime spins evolve in such a way that the coplanarity
of the spins inside the phases is preserved.
Actually, the domains compete among themselves
and grow much in the same way as the equilibrium
phases in usual coarsening systems, as testified
bye the increase of $L_V(t)$ (see Fig.~\ref{lun_piani}).
This regime lasts until the phenomenon of the breakdown of the
plains, discussed in Sec.~\ref{brakedown}, occurs (when this happens
$L_V(t)$ stops growing and goes abruptly to zero. The
end of this first dynamical stage can then
be easily recognized by inspection of Fig.~\ref{lun_piani}).

As we will explain below, in this regime two mechanisms
are at work: The former is responsible for the coarsening
of the domains, the latter is the phase-ordering
of the spins inside the domains. Since these mechanisms
are associated to two different growing lengths,
dynamical scaling is not obeyed, as we will show explicitly. 

\subsubsection{Coarsening of the domains}

In the limit of small $T$, among the moves discussed in Sec.~\ref{depinning},
those which produce the smallest energy increase are overwhelmingly favored. 
These are generally the moves involving the spins near the 
boundary of a plane, for instance those
on sites $i-1$, $i$ in Fig.~\ref{figure}. 
The effect of this move occurring in the $\beta$-phase, 
is the nucleation of a third phase,
denoted by $\gamma $, as it is clear considering the direction 
of the vector $\vec v_{i-1}$ in the second line (denoted as time $t+1$) of the figure
(a spurious phase on site $i-2$ is also 
generated, whose presence is, however, irrelevant). 
At this point the dynamics may proceed by
rotating the spins $\vec \sigma _{i-2},\vec \sigma _{i-1}$ in order to make
$\vec \sigma _{i-2}$, $\vec \sigma _{i-1}$, and $\vec \sigma _i$ coplanar
(third line, denoted as time $t+2$, in the figure). In this way the new phase 
$\gamma $ may spread replacing the pre-existing phase $\beta$.
After the complete replacement of the phase $\beta $ with $\gamma$
(when this occurs), if the energy of the system is increased (because 
$\gamma $ is more {\it distant} to the neighboring phase
$\alpha $ (or the one on the left, not shown in the figure) 
than the original $\beta$-phase,
the $\gamma $ phase is quickly re-adsorbed by reversing the process. 
Conversely, if the energy of the system is decreased
a new blocked state is reached, characterized by domains of more {\it nearby} phases. 
This mechanism provides a {\it direction} to the process, favoring on the average 
the formation of new phases for which a diminishing of the system energy occurs. 
At this point the process can start again with the activated nucleation 
of a new phase replacing $\gamma $ and so on repeatedly 
until one of the two adjacent phases (say $\alpha $) extends over
the original domain of the $\beta $ phase, increasing the typical
size of the domains.

The basic steps of this ordering process may recall 
what happens in the COP Ising model~\cite{Cornell91}. 
Actually, in both cases there are domains of 
different phases ($\sigma _i=\pm 1$ in Ising, $\vec v _i=\alpha ,\beta$ 
etc. in the Heisenberg model) separated by
sharp interfaces (see also the discussion on Porod tails below in this section).
In both cases, the first step is the nucleation of a germ of another phase 
(the evaporation of a monomer in Ising) inside a domain of a preexistent phase. 
After nucleation, the kinetics proceeds by
a random motion of the nucleated phase. This analogy is not only qualitative, 
since, as we show below, the growth exponent of $L_V(t)$ is the same ($z=3$).

\begin{figure}
    \centering
    
   \rotatebox{0}{\resizebox{.85\textwidth}{!}{\includegraphics{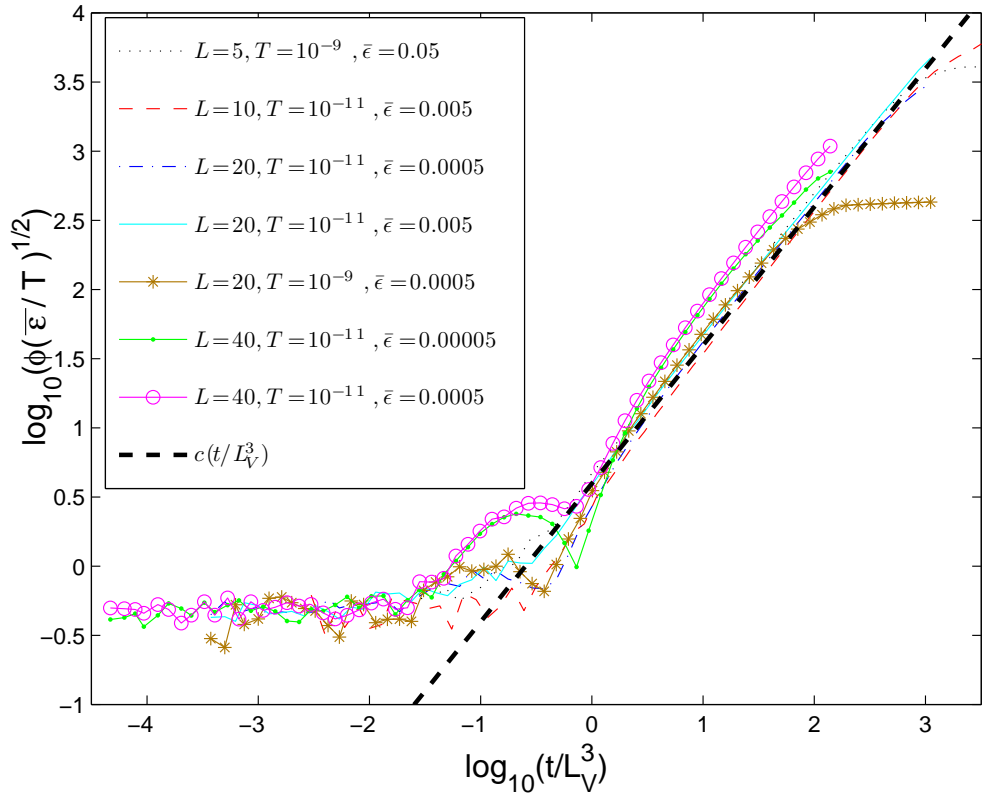}}}
\caption{Simulation of a single domain of $L_V$ spins. The angle $\phi $ of the central spin 
(rescaled by $T^{1/2}$) is plotted against rescaled time $t/L_V^3$.}
\label{angoli}
\end{figure}

Actually this can be inferred by the following argument: 
After the first move, the position of the boundary between the
new nucleated phase $\gamma $ and the remaining of the old
$\beta $ phase performs a random walk. Considering
the long time regime, where $L_V(t)$ is large,
most of the times the boundary returns to its original position
$j$. In this case the $\gamma $ phase is re-adsorbed after the
duration $\delta \sim L_V^2(t)$  
of the random walk. The probability
that the interface moves a distance $L_V(t)$ and hence the $\beta $-phase
is eliminated is proportional to $1/L_V(t)$~\cite{Cornell91}; for this reason the
whole process must be repeated $L_V(t)$ times in order to substitute
the old phase $\beta $. This requires a time
\be
\tau _{\dag}\propto L_V (t)\delta \propto L_V(t)^3. 
\label{firststep}
\ee
This conjecture has been tested by means of numerical simulations, 
mimicking the evolution of a single domain as follows: We have prepared 
a domain of $L_V$ initially coplanar spins, with a uniform rotation
such that $\epsilon _i=\overline \epsilon$, $\forall i$ in the bulk
of the domain. The boundary condition is
made of two spins (on each side) lying on a differently oriented
plane. Then we started the dynamics and recorded the behavior
of the central spin as a function of time, for different
choices of $L_V$ and of $\overline \epsilon $. The results are presented 
in Fig.~\ref{angoli}. The central spin can be described by a couple 
of angles $\phi,\psi$. Here we plot the behavior of $\phi$ 
(similar results are obtained for $\psi $) as time passes.
In an early stage $t<\tau _{\dag}$ the central spin remains blocked,
since the dynamics starts from the boundary
and proceeds towards the interior, as discussed above. Then $\phi $
is constant.
Later, from $t=\tau _{\dag}$ onwards the central spin begins to
rotate unless all the plane is aligned with the plane described 
by the boundary conditions. This is testified by the linear increase of
$\phi $ towards a limiting final value. The figure shows that the curves for
different $L_V$ collapse when time is rescaled
as $t/L_V^3$. This supports Eq.~(\ref{firststep}). Notice also that,
by considering different temperatures and $\overline \epsilon $, 
one obtains data collapse
by plotting $\phi (\overline \epsilon/T)^{1/2}$. This is a consequence
of Eq.~(\ref{typictheta}). 
Actually $d_i$ can be written as $d_i=\sqrt{(2\epsilon_i)}$.
Then $\overline d$ is of order
$\sqrt {\overline \epsilon (t)}$. 
Analogously, it can be shown that also 
$\overline p\sim \sqrt {\overline \epsilon (t)}$.
Inserting these typical quantities in Eq.~(\ref{typictheta}), and letting
$\cos \overline \theta \simeq 1-\overline \theta ^2$, since in the late
stage spins are rather aligned, one obtains
\be
\overline \theta \sim \left( \frac {T}{\overline \epsilon} 
\right)^{\frac {1}{2}}.
\label{tipth}
\ee
In conclusion, the numerical simulation of the evolution of the single
domain, confirms our scaling hypothesis.

At low temperatures, as discussed above, $\overline \theta $ is small and hence
the new generated $\gamma $ phase is only slightly different from 
the pre-existing $\beta $. 
Reasoning along the same lines as regarding the simulation of a single
domain, the process of replacement
of an old phase by a new one must be repeated a number $n_r(T)\propto T^{-1/2}$ 
of times in order to obtain, in place of the original $\beta $ phase, the
phase (say $\alpha $) of one of the neighboring 
domains. 
In conclusion, the complete replacement of the phase $\beta $ 
with a neighboring one ($\alpha $) requires a time of order 
$\Delta t \propto n_r(T) L_V^3 $.
When this process is completed the typical length of a 
domain is increased of a quantity $\Delta L_V(t)\propto L_V(t)$.
Therefore $dL_V(t)/dt\simeq \Delta L_V(t)/\Delta t\propto L_V(t)^{-2}T^{1/2}$  
and hence 
\be
L_V(t)\sim (tT^{\frac{1}{2}})^{1/3}
\label{lvv}
\ee
This prediction can be checked in Fig.~\ref{lun_piani}. 
Here one observes that the curves for $L_V(t)$ relative to different 
temperature quenches 
collapse (after the pinning) when plotted
against $tT^{1/2}$. The collapse is good for the lower temperatures
(for $T\le 10^{-4}$), while it is quite rough at higher temperatures. 
This is  expected since
our results are valid in the $T\to 0$ limit. Regarding the power 
growth law~(\ref{lvv}), 
it is satisfactorily confirmed by the data in a certain time window after the depinning.
For longer times $L_V(t)$ goes 
abruptly to zero due to the phenomenon of the breakdown of the planes, 
that will be discussed in 
Sec.~\ref{brakedown}. As explained in Sec.~\ref{brakedown} this phenomenon is delayed 
lowering $T$ (actually, for a relatively high temperature as 
$T\geq 10^{-3}$ it prevents the observation of 
the law~(\ref{lvv})). 

\begin{figure}
    \centering
    
   \rotatebox{0}{\resizebox{.85\textwidth}{!}{\includegraphics{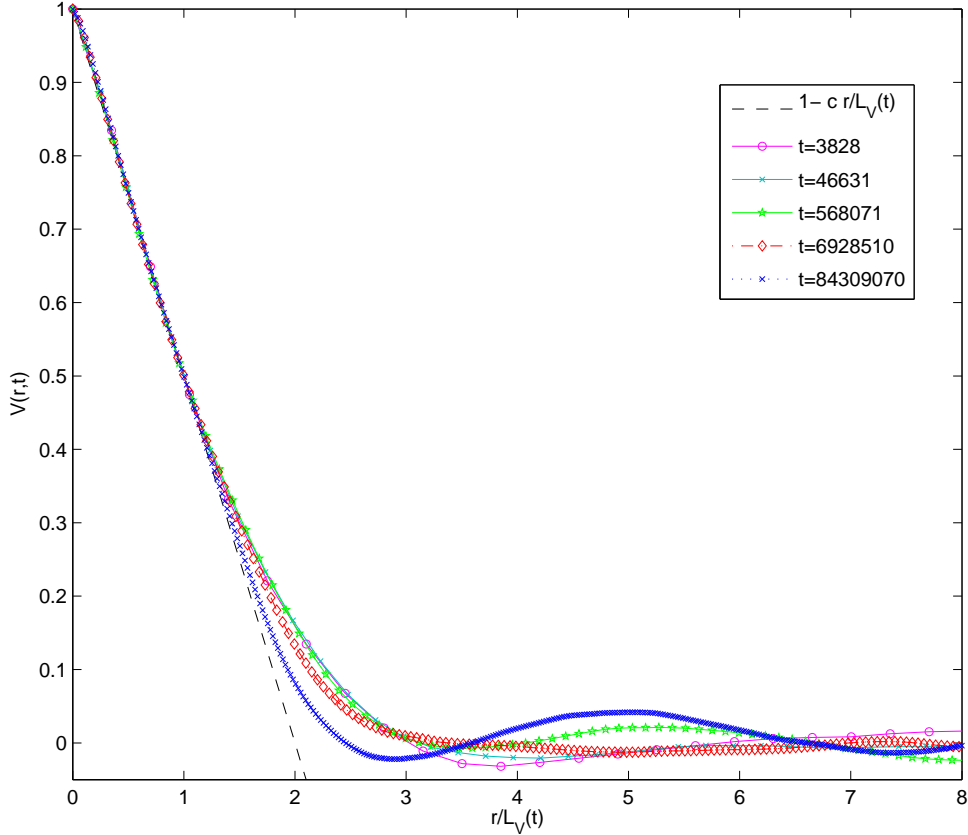}}}
    \caption{$V(r,t)$ is plotted against $r/L_V(t)$ for a quench to $T=10^{-7}$ and different times. 
The dashed line is the Porod law~(\ref{porod}).}
\label{scal_piani}
\end{figure}

We consider now the issue of dynamical scaling.
In Fig.~\ref{scal_piani} we plot $V(r,t)$
against $x=r/L_V(t)$ in the range of times in which this first dynamical regime occurs. One observes a good data collapse up to
$r/L_V(t)\simeq 1.5$
According to Eq.~(\ref{vscal}), this implies
that $V(r,t)$ takes a scaling form in this range of $r/L_V(t)$. 
Since usually scaling is first achieved for smaller distances, one 
could infer that, by pushing the simulations to much longer times,
one could observe collapse on a larger range of $/L_V(t)$ and
conclude that the whole $V(r,t)$ scales.
This is surprising, since we
have anticipated  
that dynamical scaling is violated in this regime. However 
this happens because the correlator $V(r,t)$,
due to its construction, exclusively probes the
dynamics of the boundaries of the domains, being blind
with respect to the spin configuration inside,
whose evolution is responsible for the breakdown of dynamical scaling,
as we will discuss in Sec.~\ref{ph}. One could say that, restricting
the attention on the plane boundaries, {\it scaling is obeyed},
although globally it is not. A similar situation is observed in
the $d=1$ XY model~\cite{Rutenberg95} where again scaling does not hold
(for a different reason) but particular correlators, such as $C(r,t)$ or $V(r,t)$ take
scaling forms. Clearly this is not a general property of every
observable, as genuine scaling should imply.
Finally, the Porod law~(\ref{porod}) is obeyed, signaling 
that interfaces are sharp and that domains remain well defined
while coarsening in this whole regime. 

\subsubsection{Phase-ordering of the spins} \label{ph}

\begin{figure}
    \centering
    
   \rotatebox{0}{\resizebox{.85\textwidth}{!}{\includegraphics{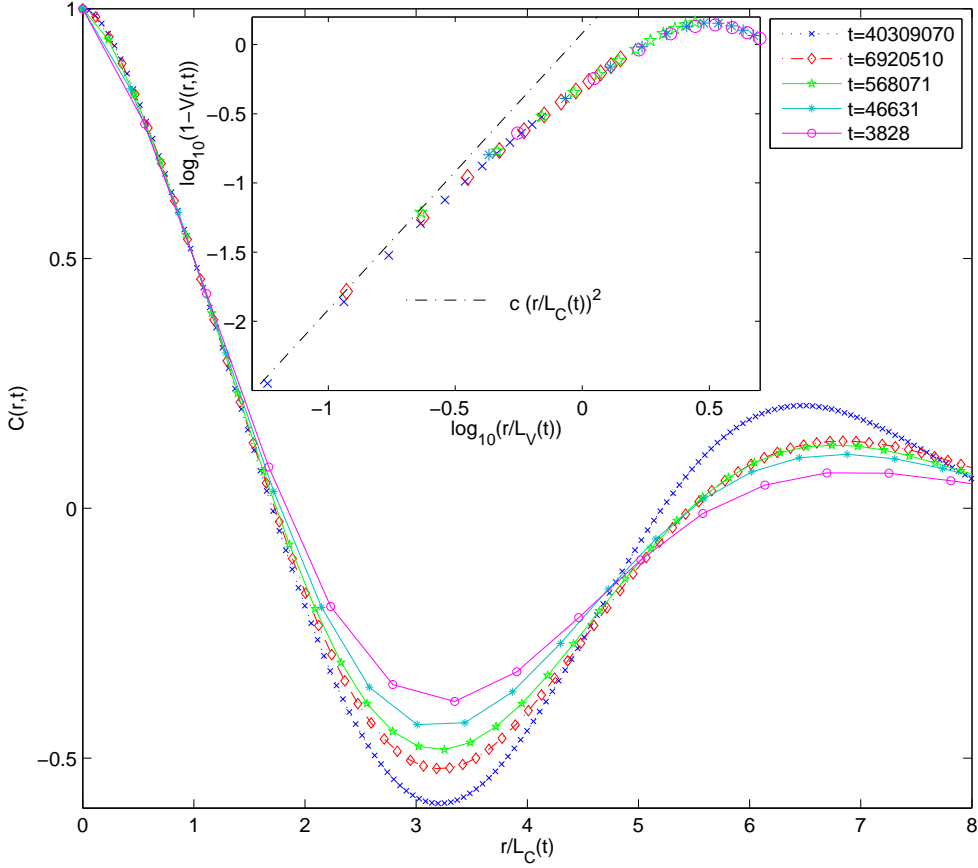}}}
    \caption{$C(r,t)$ is plotted against $r/L_C(t)$ for a quench to $T=10^{-7}$ and different times. The dashed line is the Porod law~(\ref{porod}). In the
inset, $1-C(r,t)$ is plotted against $r/L_C(t)$ on a double logarithmic
scale. The dot-dashed line is the form~(\ref{brayputo}), with $N=2$.
}
\label{scal_spin}
\end{figure}

Inside the planes, the spins smoothly rotate generating textures,
much like in the one dimensional XY model~\cite{Rutenberg95}.
Comparing Fig.~\ref{lun_spin} and Fig.~\ref{lun_piani} one understands that 
the length $L_C(t)$ associated
to the spin-spin coherence is much smaller then $L_V(t)$, particularly
for large times.
Then, what really matters for $L_C(t)$ is the evolution of
the spins in the bulk of the domains, where, 
since the interfaces are far away, all the additional complications
related to their presence become irrelevant.
As far as the spin spin correlations are considered, therefore, 
one expects the system to behave as a {\it normal} conserved vectorial
system (but with $N=2$!): Namely, $C(r,t)$ to obey the scaling form~(\ref{scalgferro}) 
and Eq.~(\ref{lc}) to hold. These features can be checked in 
Figs.~\ref{lun_spin},\ref{scal_spin}. Regarding the growth law of
$L_C(t)$ we obtain a behavior in good agreement with what expected,
namely Eq.~(\ref{lc}), for the higher temperatures, namely for 
$T\geq 10^{-4}$. For the lower temperatures, namely  
$T\leq 10^{-7}$, we measure an effective exponent
somewhat larger than $1/4$. The case with $T=10^{-6}$ is somehow in between,
since the curve initially (after the pinning) grows with an exponent
larger than $1/4$ but then the slope is gradually reduced and an exponent in
agreement with $1/4$ is obtained towards the end of the simulation. 
The behavior of this curve may probably provide an interpretation for
what observed for the lower temperatures. Namely, the 
behavior seems to set in after a transient which widens as $T$ is lowered.
In the transient a slightly larger exponent is observed.
Notice that the curves for 
$L_C(t)$ roughly collapse (we recall that the figure covers 10 decades in $T$) 
when plotted against $tT^{1/2}$, for the same
reason of $L_V(t)$. Similarly, the collapse seem to improve in quality
as  $T\to 0$, as expected, while for $T\ge 10^{-3}$ the collapse is poor.

The scaling form~(\ref{scalgferro}) can be checked in Fig.~\ref{scal_spin}.
Here one observes a good data collapse up to $r/L_C(t)\simeq 1.5$,
similarly to what observed for $V(r,t)$.
Concerning the shape of $C(r,t)$, differently from $V(r,t)$ it
does not display the Porod´s tail, as expected since
the rotation of the spins is smooth and there are no sharp
interfaces. On the other hand, one observes (in the inset) the behavior~(\ref{brayputo})
typical of vectorial systems but, interestingly, with an {\it effective}
value $N=2$ which is clearly interpreted as due to the fact that spins
in this regime lie on planes.

The results discussed insofar provide a picture of a
system with two different ordering mechanisms at work, 
coarsening of the planes and phase-ordering of the spins.
These profoundly different processes 
coexist in this regime, apparently in a rather independent
way, without interfering, possibly because they act on
different length-scales. 
To each mechanism a particular correlation function 
is naturally associated, giving rise to two distinct lengths
growing with different exponents. Due to that, dynamical scaling
is violated even if $V(r,t)$ and $C(r,t)$ possibly scale separately
with respect to $L_V(t)$ and $L_C(t)$.

\subsection{Breakdown of the domains} \label{brakedown}

As discussed above, although thermal fluctuations become relevant in
depinning the system, their effect in the previous regimes
is basically to produce the coarsening of the planes without
dissolving them. This is because, 
since in the early stage the spins are quite misaligned,
both $\overline d$ and $\overline p$ are rather large and hence 
according to Eq.~(\ref{typictheta}), the typical rotation angle
$\overline \theta $ is rather small.

As the dynamics proceeds, however, textures stretch, spins align,
and $\overline d$ and $\overline p$ decrease. 
In view of Eq.~(\ref{typictheta}), at a certain time 
$\tau _v(T)$, values of $\overline \theta $ sufficiently large,
namely $\cos \overline \theta  -1$ of order unity, become available, which are 
sufficient to destroy the structure of the domains.
The breakdown of the planes can be nicely detected by
directly looking at the spin configuration or, more
properly, by inspection of $V(r,t)$. In fact, while for $t <\tau _v(T)$, 
$V(r,t)$ takes the scaling form~(\ref{vscal}),
for $t>\tau _v(T)$, when the domain disappear, it 
quickly collapses to a rapidly decaying function. In the meanwhile,
$L_V(t)$ stops growing and abruptly decreases,
as shown in Fig.~\ref{lun_piani}.

\begin{figure}
    \centering
    
   \rotatebox{0}{\resizebox{.85\textwidth}{!}{\includegraphics{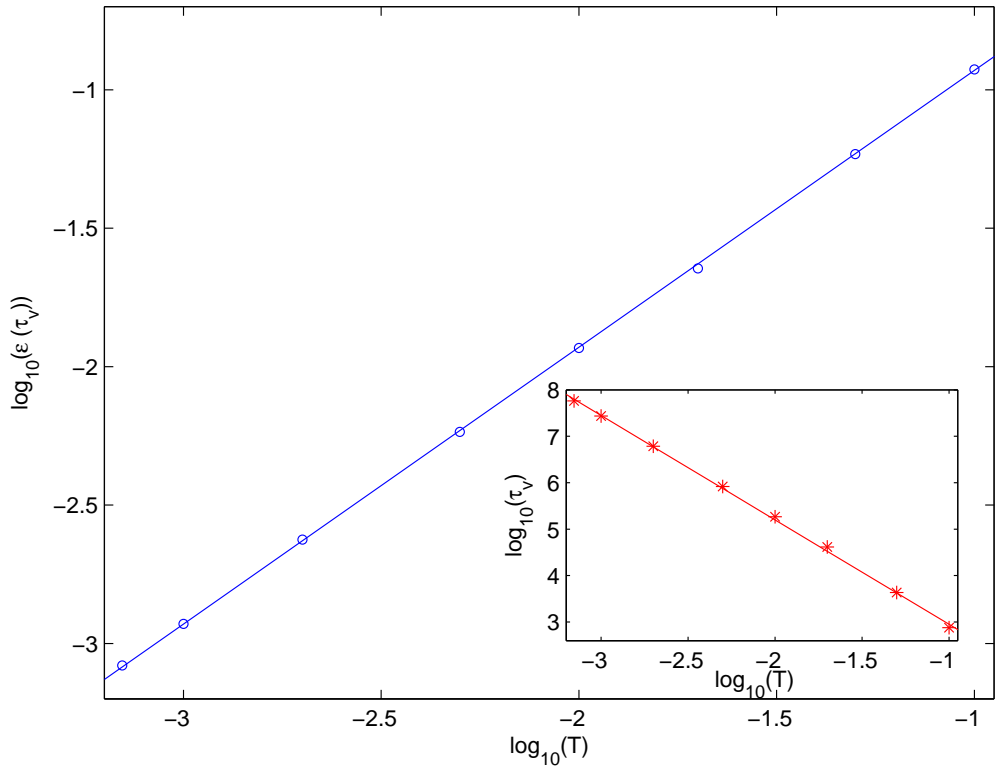}}}
    \caption{The quantity $\overline \epsilon (\tau _v)$ is plotted against
     $T$. The continuous line is the best fit 
     $\overline \epsilon (\tau _v)=AT$, with $A=2.76$. In the inset
     $\tau _v(T)$ is plotted against $T$. The continuous line is the best
     Power-law fit $\tau _v (T)=KT^{-2.3}$.}
\label{crolli}
\end{figure}

$\tau _v(T)$ can be evaluated by means of Eq.~(\ref{tipth}):
The condition $\cos \overline \theta  -1\sim 1$
for the breakdown of the plains is realized when
\be
\overline \epsilon (\tau _v)\simeq AT,
\label{Ecrollo}
\ee 
where $A$ is a constant.
This is very well confirmed numerically: In
Fig.~\ref{crolli}, we plot the average energy computed at the time
when $L_V(t)$ reaches its maximum, against the temperature of the quench.
We find that Eq.~(\ref{Ecrollo}) is well verified with $A\simeq 2.76$.

In order to estimate $\tau _v(T)$ 
we notice that $\overline \epsilon(t)$ is entirely associated with the 
smooth rotation
of the spins in the bulk of the domains, because
on the interface between domains spins are
perfectly aligned. The typical angle between two adjacent spins
is $\overline \phi \sim L_C^{-1}(t)$. Hence the energy 
$\overline \epsilon(t)= 1-\cos \overline \phi \sim L_C^{-2}(t)$,
a fact that we have verified with good accuracy in the simulations. This leads to 
\be
\tau _v(T)\simeq KT^{-\chi} \quad ,\quad \chi=\frac{5}{2},
\label{tauv}
\ee
where $K$ is a constant.
The dependence of $\tau _v(T)$ on temperature is shown in the inset of
Fig.~\ref{crolli}. We find a power-law behavior but with a value  
$\chi \simeq 2.3$ only in rough agreement with the expected
value $\chi =5/2$. This partial agreement is probably due to
the fact that our results are only valid in the asymptotic limit
of small $T$ and large times $t$. Here, for instance, we cannot 
consider very small temperatures since for $T<10^{-3}$ the time of domain breakdown 
is too long for our simulations.
However, as already observed, in such large temperature regime
the numerical data do not scale according to the asymptotic behavior 
(recall the discussion regarding Figs~\ref{lun_spin},\ref{lun_piani}
in Sec.~\ref{regime1}).  

Since the time to approach and leave the frozen state is negligible
in the $T\to 0$ limit, $\tau _v(T)$ represents also the duration
of the regime where scaling is violated.
Notice that it
increases quite rapidly as quenches are made deeper.

\subsection{Second phase-ordering regime: absence of domains}
\label{regime2}

We have seen that the breakdown of the domains occurs when
the energy $\overline \epsilon(t)$ of the system is comparable 
to $AT$ ($A\simeq 2.76$), Eq.~(\ref{Ecrollo}).
On the other hand,
the system equilibrates at the time $\tau _{eq}(T)$ when
$\overline \epsilon(t)$ reaches the equilibrium value 
$\overline \epsilon(\tau _{eq})=E_{eq}(T)\simeq T$.
Hence the energy must still be lowered after the breakdown of the
planes. Then, phase-ordering must continue even after $\tau _v(T)$. 
Clearly,
once the domains are eliminated, smooth rotations of the
order parameter remain the only mechanism at work, and one expects the usual
coarsening mechanism of vectorial systems for $L_C(t)$, characterized by
dynamical scaling with $z=4$. 
This can be checked in Fig.~\ref{lun_spin}. One can observe that
the power law~(\ref{lc}) continues to be valid, with no apparent
modifications, even after the breakdown of the planes, 
signaled by the decrease of $L_V(t)$ (Fig.~(\ref{lun_piani})).

\subsection{Equilibration} \label{equilib}

Recalling that
$E_{eq}(T)\simeq T$,
using again $\overline \epsilon (t)\sim L_C(t)^{-2}$
one obtains 
\be
\tau _{eq}(T)\simeq A^2 \tau _v(T).
\label{taueq}
\ee
This results shows that the duration of this regime characterized by 
dynamical scaling is comparable to that of the previous one.
Their duration diverges with the same exponent as $T\to 0$.

In the simulations, after $\tau _{eq}(T)$ the system is observed 
to enter the equilibrium stationary state. Computing the behavior
of some equilibrium quantities, as, for instance, $E_{eq}(T)$
or $\xi (T)$, we found the exact equilibrium results of Sec. \ref{model}
with great accuracy. This confirms the correctness and the efficiency
of the heat bath transition rates \ref{heatb1}. 

\section{Summary and conclusions} \label{concl}

In this paper we have studied the kinetics of the one-dimensional
Heisenberg model with conserved order parameter. 
The distinguishing feature of this model
is the presence of defects in the form of
couples of parallel spins separating coplanar regions.
These are quite unusual and somewhat counterintuitive defects,
since normally one associates the notion of defect to regions where
the order parameter varies quite abruptly, while here
spins are perfectly aligned on the defect. Their nature, however, is
clearly manifest in the representation of the $\vec v_i$,
where they qualify as unstable (but long living in deep quenches),
point like defects. Their presence makes the kinetics similar
in some respect to that of a scalar order parameter, because defects
play the role of interfaces in the $\vec v_i$-representation.
In particular, since the removal of defects can only be achieved
by activated moves, in a low temperature quench the system initially
pins; the later thermally-activated evolution is characterized by
coarsening of the domains with the {\it scalar}-like exponent $z=3$.
The vectorial nature of the system makes itself manifest
particularly in the smooth rotations of the spins inside the domains,
producing alignment over a typical length growing with
an exponent $z=4$ characteristic of vectorial systems.
This interplay between two different ordering mechanisms
continues up to a time $\tau _v (T)$ (Eq.~(\ref{tauv})),
which represents the lifetime of unstable defects. After,
defects are removed by thermal fluctuations and a second
phase-ordering regime sets in, characterized only by smooths
variations of the spins, where dynamical scaling is obeyed.
It is worth mentioning, that the duration of the second
phase-ordering regime, without defects, is comparable
with respect to that of the first one in the low temperature
limit.    

These features are unusual and unexpected in non-disordered phase-ordering
systems. A natural question, therefore, regards their generality, namely
if one could expect a similar behavior in other systems. 
The peculiar dynamics found in the Heisenberg chain is obviously related
to the conserved character of the kinetics. Therefore we do not
expect to find something similar in systems without the conservation law,
because in that case parallel spins can be singularly updated and, in
doing so, the defect is removed. In order to check this we have
performed simulations of the system subjected to the same Hamiltonian
but with a dynamics which does not conserve the magnetization.
As expected, in this case we did not find the unusual features
observed with COP, such as scaling violations. The same is found
by considering a dynamics where conservation is imposed only
globally (by exchanging two spins without the constraint of 
neighborhood), as expected since it is known \cite{Bray94,nonlocal}
that the global conservation law is irrelevant. 
Restricting to systems where conservation is realized locally, 
as far as we can see there is
not a reason preventing the formation of similar defects for
$N>3$ or, perhaps, even with $d>1$. The other ingredient which
turns out to be fundamental in the model considered insofar is
the ability of the system to squeeze all the $n$ spins between two
defects into a plane, during the regime preceding the pinning.
This reduction of the effective internal dimensionality of the
order parameter from $N$ to $N-1$ can be achieved because, since
$n$ is finite, these spin can be projected on the most energetically
favorable plane in a finite time (the time over which the system pins
at $T=0$). On lattices, the feature of $n$ being finite is related to
the one-dimensional nature of the system. In fact, in $d>1$ the typical
configuration of the system is a bi-continuous percolating structure~\cite{Bray94}
and, even assuming that a relevant number of couples of parallel spins may be formed in a early stage, their geometry should not enclose a domain with
a finite number $n$ of spins. Actually, we have run some simulation of the 
Heisenberg model in $d=2$ and we have not find the peculiarities
of the one-dimensional case. On inhomogeneous systems, $n$ is expected to be 
finite on finitely ramified structures. Therefore, a similar behavior could be observed 
on comb lattices, t-fractals and other finitely ramified networks, where phase ordering 
for discrete models display 1-dimensional features \cite{PRL,PRE}.

On the basis of these reasoning, we also infer
that a behavior similar to that of Heisenberg chain with COP could
be expected for a generic ${\cal O}(N)$ model (with $N\geq 3$) with
COP in $d=1$ or, possibly, on finitely ramified networks. Let us mention that our simulations of the ${\cal O}(4)$ model 
quenched to $T=0$ show pinning in states characterized by defects
similar to those of the Heisenberg model. A rather complete analysis of the
kinetics for generic $N$, similar to that presented in this Article, is geometrically rather complicated and beyond the scope of this paper,
but may represent an interesting issue for further research.   

The authors thank Francesco Di Renzo for useful discussions about numerical implementations of the heat-bath 
algorithm.


\begin{thebibliography}{99}
\bibitem{notadin}
For instance, in the case of the XY model with non conserved dynamics,
we have checked that with the transition rates~(\ref{heatb1}) 
the asymptotic analytic expressions of Ref.~\cite{Rutenberg95} can be 
verified numerically with great accuracy, while using other transition rates, 
since the dynamics is incomparably slower, the asymptotic domain is 
not accessed in any reasonable simulation time.
\bibitem{notacrit}
Let us notice that in this case $T=0$ is also the critical
temperature of the model. However the kinetics in this
case displays all the phenomenology of phase-ordering,
and not the radically different one of critical dynamics,
since the final state towards which the system evolves is
completely ordered. A careful discussion of this point
is contained in \cite{teff}.
\end{thebibliography}
\end{document}